      [1999/12/01 v1.4c Il Nuovo Cimento]
\documentclass{cimento}

\usepackage{graphicx}  
\title{ The CODEX-ESPRESSO experiment: cosmic dynamics, fundamental physics, planets and much more...}
\author{S.~Cristiani\from{ins:1}\from{ins:2}\ETC,
G.~Avila\from{ins:3},
P.~Bonifacio\from{ins:1}\from{ins:13},
F.~Bouchy\from{ins:9},
B.~Carswell\from{ins:5},
S.~D'Odorico\from{ins:3},
V.~D'Odorico\from{ins:1},
B.~Delabre\from{ins:3},
H.~Dekker\from{ins:3}
M.~Dessauges\from{ins:4},
P.~Dimarcantonio\from{ins:1},
R.~Garcia-Lopez\from{ins:6},
A.~Grazian\from{ins:8},
M.~Haehnelt\from{ins:5},
J.~M.~Herreros\from{ins:6},
G.~Israelian\from{ins:6},
S.~Levshakov\from{ins:10},
J.~Liske\from{ins:3},
C.~Lovis\from{ins:4},
A.~Manescau\from{ins:3},
E.~Martin\from{ins:6},
M.~Mayor\from{ins:4},
D.~Megevand\from{ins:4},
P.~Molaro\from{ins:1},
M.~Murphy\from{ins:5}\from{ins:12},
L.~Pasquini\from{ins:3},
F.~Pepe\from{ins:4},
J.~Perez\from{ins:6},
D.~Queloz\from{ins:4},
R.~Rebolo\from{ins:6},
P.~Santin\from{ins:1},
P.~Shaver\from{ins:3},
P.~Span\`o\from{ins:7},
S.~Udry\from{ins:4},
E.~Vanzella\from{ins:1},
M.~Viel\from{ins:1},
M.~R.~Zapatero\from{ins:6},
F.~Zerbi\from{ins:7},
S.~Zucker\from{ins:11}}

\instlist{\inst{ins:1} INAF/National Institue for Astrophysics, Osservatorio di Trieste, Italy
\inst{ins:2} INFN/National Institute for Nuclear Physics, Trieste, Italy
\inst{ins:3} European Southern Observatory, Garching, Germany
\inst{ins:4} Observatoire de Gen\`eve, Switzerland
\inst{ins:5} Institute of Astronomy, University of Cambridge, Cambridge, UK
\inst{ins:6} Instituto de Astrofisica de Canarias, Spain
\inst{ins:7} INAF/National Institue for Astrophysics, Osservatorio di Merate, Italia
\inst{ins:8} INAF/National Institue for Astrophysics, Osservatorio di Roma, Italia
\inst{ins:9} Institut d'Astrophysique de Paris, France
\inst{ins:10} Ioffe Physico-Technical Institute, St.Petersburg, Russia
\inst{ins:11} Dept. of Geophysics \& Planetary Sciences,
Tel Aviv University, Israel
\inst{ins:12} Centre for Astrophysics \& Supercomputing, Swinburne
University of Technology, Melbourne, Australia
\inst{ins:13} GEPI, Observatoire de Paris, CNRS, Universit\'e Paris
Diderot, Meudon, France }
\begin{document}
\maketitle
\begin{abstract}
CODEX, a high resolution, super-stable spectrograph to be fed by
the E-ELT, the most powerful telescope ever conceived, will for
the first time provide the possibility of directly measuring the
change of the expansion rate of the Universe with time and much more,
from the variability of fundamental constants to the search for other
earths. A study for the implementation at the VLT of a precursor of
CODEX, dubbed ESPRESSO, is presently carried out by a collaboration
including ESO, IAC, INAF, IoA Cambridge and Observatoire de Gen\`eve.
The present talk is focused on the cosmological aspects of the
experiment.
\end{abstract}

\section{Introduction}
Hubble's discovery of the expansion of the Universe, enabled in the late 20s
by the new observational facilities on Mount Wilson,
brought to an end the cherished belief held by most physicists of the time,
including Albert Einstein, that the Universe is static and not evolving.

The discovery has later been impressively confirmed by a vast range of astronomical
observations and led to the now widely accepted Hot Big Bang Theory,
which rests on firm pillars such as the detection of the relic Cosmic Microwave Background
(CMB,~\cite{CMB1,CMB2})
and the experimental verification of the prediction for the
synthesis of light elements~\cite{BBN}.

The Hot Big Bang is now an essential aspect of the cosmological standard model
and the central question has become:
{\it what is the stress-energy tensor of the Universe?},
which - with the ansatz of homogeneity and isotropy - reduces to:
{\it what is the mean density and the equation of state
of each mass-energy component of the Universe?}
Since these parameters determine both the evolution with time
and the geometry of the metric that solves the Einstein equation
one can use a measurement of either to infer their values.

The above question has been addressed by various observations:
from CMB and SN Type Ia
to the abundance of clusters, from weak lensing
to the large scale structure traced by galaxies and the inter-galactic
medium (IGM),
and the results are consistent with the so-called {\it Concordance Model},
a Friedmann-Robertson-Walker Universe with
no curvature, which has provided a first moderate embarrassment,
due to a significant component of dark matter whose nature remains elusive, and
a major surprise with the discovery that the expansion of the Universe
has recently begun accelerating for physical reasons basically unclear
at the moment.
The latter is accommodated by modifying the stress-energy
tensor to include a new component with negative pressure, something
reminiscent of the old ether, falsified by the experiment of Michelson
and Morley.

It is important to note that the above mentioned experiments (CMB, Lensing,
SNIa, etc.) are based on geometry, some of them require a prior on spatial
curvature and a detailed understanding of the growth of density
perturbations, hence a specific cosmological model. For this reason,
measuring the dynamics of the Universe in a clean a direct way would
be a new fundamental test of General Relativity and Cosmology.

\begin{figure}
\includegraphics[scale=0.5,angle=270]{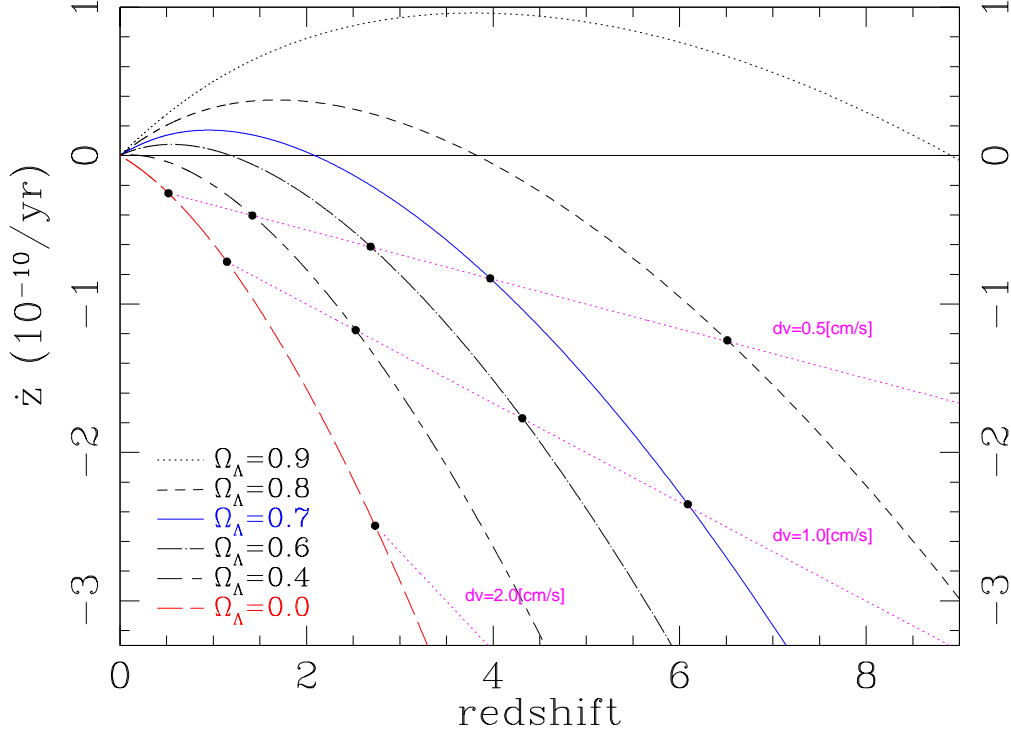}
\caption{Evolution of $\dot{z}$
as a function of redshift. The cosmological parameters have
been fixed to $\Omega_{tot}=1$, $H_{0}=70$ and different values of
$\Omega_{\Lambda}$ have been considered.
The lines represent the behavior of the Eq.~\ref{eqn:zdot}, assuming
the observation of the Lyman-$\alpha$ line.}
\end{figure}

\section{Probing the nature of the universal expansion}
Is it possible to directly measure the history of the expansion?
The goal is to reconstruct the evolution of $a(t)$,
the scale factor as a function of time.
We ordinarily measure $a(z)$ and to recover the unknown $a(t)$ we need to know
$da/dt(z)$.
In other words we need to measure the Hubble constant, $H(z) \equiv
\dot a / a$ using dynamics.

A way to implement this experiment is
to monitor the redshifts of cosmological sources.
The change of these redshifts as a function of the observer's
time is a direct signal of the acceleration of universe's expansion
and hence of its dynamics.

The variation of the redshift as a function of time~\cite{LOEB,CODEX1},
turns out to be a simple difference between the Hubble constant at the present epoch
and the one at the time of the emission
\begin{equation}
\dot{z} = (1+z)~H_{o} - H(t_{e}).
\label{eqn:zdot}
\end{equation}
The practical problem is the smallness of the signal.
Fig. 1 shows what is expected for a concordance-like model:
we are dealing with accelerations of the order of
$1$ cm/s/yr!
The signal is small but displays a characteristic signature: a pattern
of shrinking of the spectrum at high-z and stretching at low-z.
This is typical of a non-zero
cosmological constant: the zero crossing depends only on the ratio $\Omega_{\Lambda}/
\Omega_m$  ($z_o\simeq 2$ in a Concordance
Cosmology) and the amplitude on $H_o$.
\begin{figure}
\includegraphics[scale=0.5,angle=270]{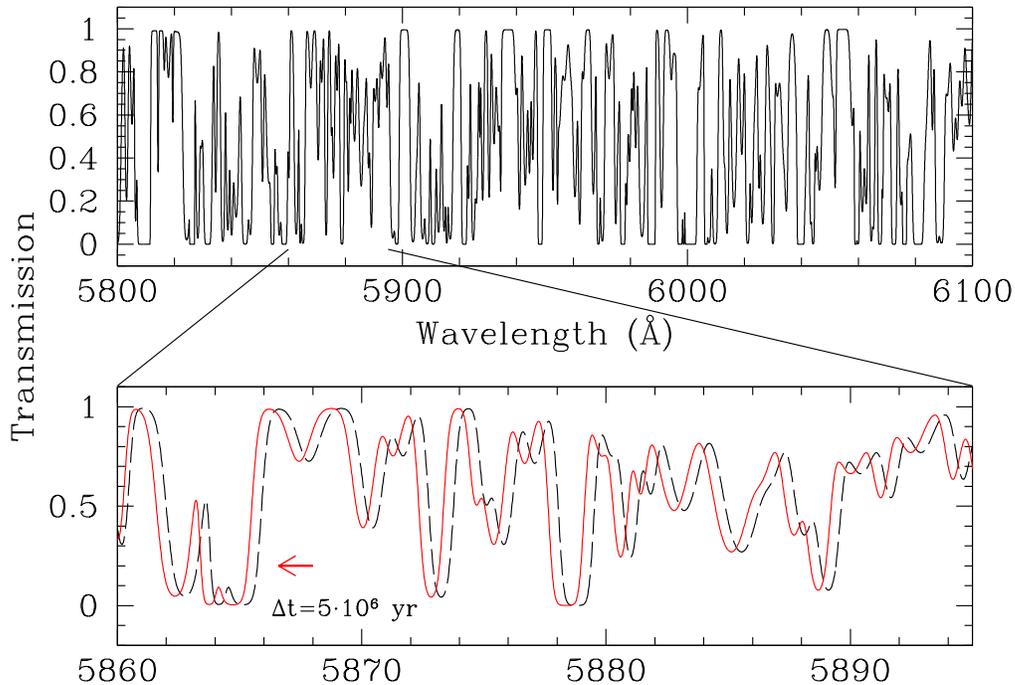}     
\caption{Upper panel: lhe transmission of the intergalactic medium at redshift
$z\sim 4$, observed along the line of sight to the QSO Q0000-26.
The average transmission is about $0.5$ and hundreds of features
modulate the signal between 0 and 1. Lower panel: the expected shift
in the spectral features due to the cosmic deceleration in a time
interval of $5 \cdot 10^6$yr.}
\label{z4forest}
\end{figure}

\section{Measuring the Cosmic Signal}
How to measure this signal?
Masers (e.g.~\cite{MASER}) or molecular lines could appear very good candidates: lines are
sharp and wavelengths can be measured with very high accuracy, due to
their rather long wavelength, noise is also much less of a problem
than at optical wavelength.
However, these targets turn out not to be well suited, since
they typically reside in deep potential wells and are subject to large
peculiar accelerations, of the same order or larger than the cosmic
signal.

Absorptions from the many intervening lines in front of high-z QSOs are
the most promising candidates. Simulations, observations and analysis
all concur in indicating that the Lyman forest and associated metal lines
are produced by systems sitting in a warm IGM following beautifully
the Hubble flow~\cite{RAUCHrev}.

The idea is then to observe the Lyman forest of a number of QSOs,
uniformly distributed all over the sky,
today and few years from now, and measure the
shrinking/stretching pattern of the absorption features (see
Fig.~\ref{z4forest}).
The variation of the normalized transmission is expected to be
of the order of $10^{-6}$ in ten years.
Is it feasible?

Difficult as it may be, such an experiment is no more complex, nor more
expensive, nor of less fundamental importance than what our colleagues
at CERN regularly do.

Accuracies not far from what we need for detecting the cosmic signal
are presently being reached in the observations of radial velocity
perturbations induced by extra-solar planets (e.g. HARPS~\cite{HARPS}).
We want to do the same but with objects that are hundred thousand times
fainter than the extra-solar planets targets, and on timescales of decades.
An extremely large light bucket is needed and in this respect the E-ELT is going to
play for European astronomers
the role of LHC for particle physicists.

\section{The CODEX Team}
On the basis of the conviction enunciated above,
the CODEX (COsmic Dynamics EXperiment) team has formed~\cite{Pasquini05}.
From the work carried out by the team to explore the feasibility
of the experiment~\cite{CODEX1} we show here the results aimed
at quantifying the optimal redshift range and instrumental
characteristics.
They can be summarized by the formula:
\begin{equation}
\sigma_v = 2 { \left ( {{\rm S/N} \over {2370}} \right ) }^{-1}
{ \left ( {{\rm N_{epochs} ~ N_{QSO}} \over {60}} \right ) }^{-{{1}\over{2}}}
{ \left ( {{\rm 1+z_{QSO}} \over {5}} \right ) }^{-1.7}
 ~ {\rm cm/s }
\label{eqn:SN}
\end{equation}
where the last exponent changes to $-0.9$ at $z_{QSO}>4$, due to the
crowding of features in the high-redshift Lyman forest.
A resolution $\Re \geq 30000$ is sufficient to fully resolve the Lyman
features. A higher resolution $\Re \sim 10^5$ would however be
advisable in order to better identify metal lines in the spectra.
\begin{figure}
\includegraphics[scale=0.51,angle=270]{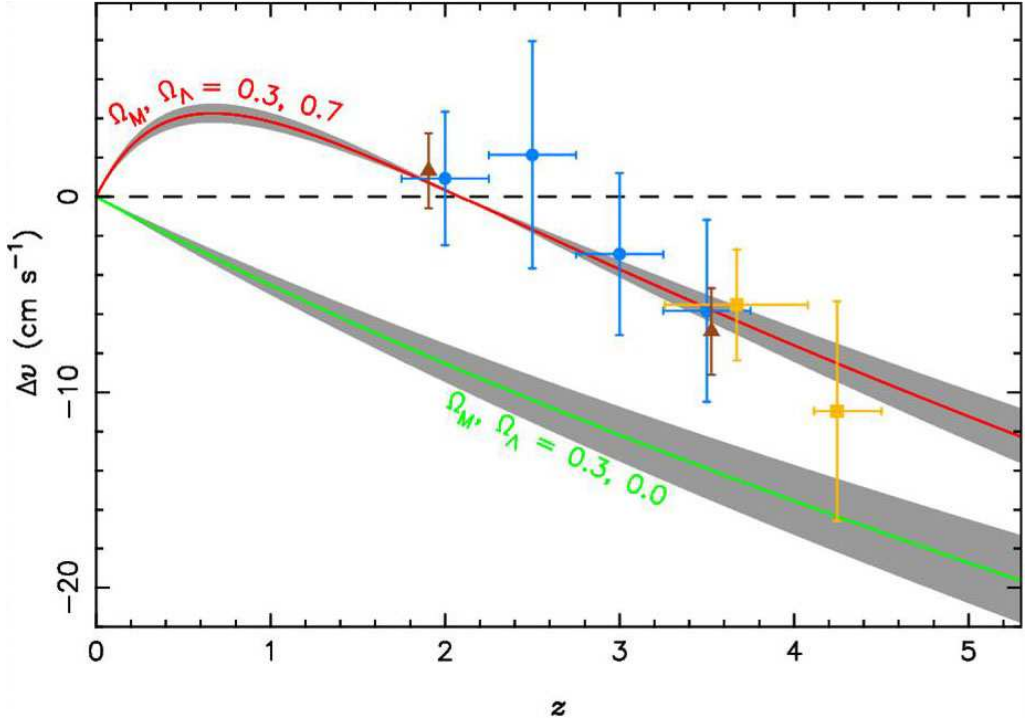}     
\caption{CODEX observing strategy assuming 2.2 nights/month of
observation with a 42m ELT over 15 years.
The three different sets of data points represent different implementations
of the redshift drift experiment, each being optimal for a different goal:
Blue points: $20$ targets (in 4 bins), selected to give the highest
overall radial velocity accuracy ($2.13$ cm/s).
Yellow points: $10$ targets selected to give the largest possible
significance of a non-zero detection. Brown points: 2 targets,
selected to give the best constraints
on the acceleration and dark energy. The grey shaded areas around the
curves correspond to the present $H_o$ uncertainty of $+/-8$ km/s/Mpc.}
\label{fig:obs}
\end{figure}
Different observing strategies are being explored, according to
different goals of the experiment (see Fig.~\ref{fig:obs}) and it is
reassuring to know that already today there are at least $20$ known
QSOs with redshift between 2 and 5 bright enough to achieve a radial velocity accuracy
of $3$ cm/s in $3200$ hours of observation with a $42$ m ELT.

Simulations have dictated the requirements for the spectrograph: a
visible, high-resolution spectrograph with an exquisite long-term
stability.
To achieve this goal with an ELT is not trivial at all and requires an
advanced design based on multiple modules (presently five are
envisaged) based on R4 echelle gratings.


A long list of potential problems is being investigated, including
secular changes in the structure of the IGM,
the variability of the QSO continuum,
weak lensing,
the heliocentric correction,
instrumental issues such as the guiding accuracy,
temperature control and
the scrambling of light in the fibers.
None at the moment is recognized to be a show-stopper.

A special issue is represented by the wavelength calibration, because
long-term stability is a must and serious doubts have been cast over
the possibility to achieve the required accuracy with conventional
lamps.
This is the reason why a new concept is being developed: the laser
frequency comb, an optical or near-IR laser generating a train of
femtosecond pulses with the pulse repetition controlled by an atomic
clock, producing a reference spectrum of evenly spaced $\delta$
functions~\cite{COMB}

Of course an instrument like CODEX would have many other applications,
among which it should be mentioned
the variation of fundamental constants~\cite{DeltaA} to a
precision of one part over $10^8$, or the detection of terrestrial extra-solar planets,
in particular following-up earth-mass candidates discovered through
other techniques, and Big Bang nucleosynthesis, with the
determination of detailed primordial abundances in order to clarify
some of the present tensions existing for example about isotopic
ratios such as $^6$Li/$^7$Li~\cite{Li6}.

Is the leap from HARPS on the ESO 3.6m telescope in La Silla to the
E-ELT too daring? Maybe.
This has prompted the CODEX Team to study the possibility of building
a precursor, dubbed ESPRESSO (Echelle Spectrograph for PREcision Super
Stable Observations) to be placed at the VLT, possibly at the
incoherently combined focus of the four UTs.
The instrument would look like one CODEX module and would allow the
community to get a first glance of a significant part of the CODEX immediate
science, of course with the exception of the cosmic dynamics.
ESPRESSO would make it possible tests of the stability of the IGM that on the
one hand are crucial for the CODEX feasibility
and on the other hand would provide fundamental
information on the formation of cosmic structures and feedback at
high-redshift (see, for example~\cite{IGMlens}).
\acknowledgments
A special thank to J.Liske, E.Vanzella and A.Grazian for their contribution to
the preparation of this article.

\end{document}